\documentclass[amsmath,
nofootinbib,
amssymb,aps,twocolumn,prl]{revtex4}

\usepackage{amssymb,graphicx}
\usepackage{amsmath,amsfonts}

\newcommand{\be}{\begin{equation}}
\newcommand{\ee}{\end{equation}}
\newcommand{\beqa}{\begin{eqnarray}}
\newcommand{\eeqa}{\end{eqnarray}}

\newcommand\m{\mu}
\newcommand\D{\Delta}
\newcommand\n{\nu}
\renewcommand\r{\rho}

\renewcommand\l{\lambda}

\def\e{{\rm e}}
\def\d{\partial}
\newcommand{\bseq}{\begin{subequations}}
\newcommand{\eseq}{\end{subequations}}

\renewcommand{\Im}{\mathop{\rm Im}\nolimits}

\newcommand{\bra}[1]{\langle #1 |}
\newcommand{\ket}[1]{| #1 \rangle}

\newcommand{\vk}{\varkappa}

\begin{document}

\begin{flushright}  
CERN-PH-TH/2014-039, 
INR-TH/2014-006
\end{flushright}
\vskip -0.9cm
\title{From scale invariance to Lorentz symmetry}

\author{Sergey Sibiryakov$^{a,b,c}$}
\affiliation{{$^a$} \it 
Theory Group, Physics Department, CERN, CH-1211 Geneva 23,
Switzerland}
\affiliation{{$^b$} \it 
FSB/ITP/LPPC, \'Ecole Polytechnique F\'ed\'erale de Lausanne,
CH-1015 Lausanne, Switzerland}
\affiliation{{$^c$}\it Institute for Nuclear Research of the
Russian Academy of Sciences, \\   
      \it  60th October Anniversary Prospect, 7a, 117312
      Moscow, Russia}

\begin{abstract}
It is shown that a unitary translationally invariant field theory
in $(1+1)$ dimensions satisfying isotropic scale invariance,
standard assumptions about the spectrum of states and operators and the
requirement that signals propagate with finite velocity
possesses an infinite dimensional symmetry given by one or a product
of several copies of conformal algebra. In particular, this implies 
presence of one or several Lorentz groups acting on the operator
algebra of the theory. 
\end{abstract}

\maketitle

Theories invariant under scaling transformations play an important role
in the
physics of fundamental interactions, condensed matter physics and
statistical mechanics.
 They appear at the fixed points of the renormalization
group (RG) flows which describe the deep ultraviolet and infrared
limits of the theory. 
We focus on the so-called ``isotropic'' scaling
transformations when the time and all space coordinates scale in the
same way, 
\be
\label{txscaling}
t\mapsto\l t~,~~~{\bf x}\mapsto \l {\bf x}\;,
\ee
where $\l$ is constant. This is the only possible form of scale
invariance in
relativistic field theories, where it is also believed to imply the 
larger conformal symmetry. This assertion has been
rigorously proved in the case of 2 space-time dimensions,
where the symmetry (\ref{txscaling}) combined with unitarity
and some technical assumptions about the energy-momentum tensor
(EMT) indeed leads to conformal invariance 
\cite{Zamolodchikov:1986gt,Polchinski:1987dy}. Generalization of this
theorem to higher space-time dimensions is a long standing
problem, see  
\cite{Luty:2012ww,Dymarsky:2013pqa} for recent progress.  

A priori, the symmetry (\ref{txscaling}) can also appear in theories
that are not relativistic. However, there
is a large amount of theoretical evidence that the RG fixed points
characterized by (\ref{txscaling}) exhibit emergent Lorentz
symmetry.
This happens in the models of particle physics with broken Lorentz invariance 
\cite{Nielsen:1978is,Chadha:1982qq,Iengo:2009ix,Giudice:2010zb,Anber:2011xf}, 
as well the models of condensed matter systems 
\cite{Vafek:2002jf,Franz:2002qy,Lee:2002qza,
Herbut:2009qb,Giuliani:2011dc,Grover:2013rc},
where an effective low-energy Lorentz symmetry (not to be confused 
with the Lorentz invariance of the fundamental forces) 
emerges from intrinsically
non-relativistic Hamiltonians.
A partial explanation of this phenomenon is provided by the general 
observation that in a unitary conformal field theory (CFT) most
of perturbations violating Lorentz invariance are 
irrelevant which implies that conformal fixed points are generically 
infrared stable \cite{Sundrum:2011ic,Bednik:2013nxa}. 
However, this argument assumes the existence of a Lorentz invariant
fixed point in the first place and does not tell anything about the
criteria for this to happen.

In this Letter I explore the relation between scale and Lorentz
invariance in $(1+1)$ dimensions. I will show that the isotropic scale
invariance\footnote{It is worth to point out the difference from
  \cite{Hofman:2011zj} which studies the consequences of {\em chiral}
  scale invariance and thus introduces from the beginning a predefined
light-cone structure.}
(\ref{txscaling}) supplemented by assumptions listed below implies that the
symmetries of the theory include one or several 
copies\footnote{An elementary example of the latter
  situation 
is provided by a
theory of several massless fields --- scalar or fermions --- 
propagating 
with  different velocities.}   of 
the full conformal algebra (which, of course,
contains the Lorentz boosts).
Namely, I am going to prove the following

{\bf Theorem}: Consider a local quantum field theory in $(1+1)$
dimensions obeying the 
symmetry (\ref{txscaling}). Lorentz
invariance is not assumed. Instead, we postulate the 
following properties:

{\bf 1.}{\it Translation invariance.} The theory is invariant under
translations in time and space which correspond to local conserved
currents. It is convenient to combine these into the EMT\footnote{The
  reader should not be confused by the relativistic notations: though
  we use the space-time indices $\m,\n$ 
  etc. that take values $t,x$, we cannot raise or lower them due to
  the absence of a 
  Lorentz metric. We will 
  sometime combine the time and space coordinates into a single vector
$\xi^\m$ with $\xi^t=t$, $\xi^x=x$. Summation over repeated indices is
assumed unless stated otherwise.} $T^\m_\n(t,x)$
whose conservation implies
$\d_\m T^\m_\n=0$.
The energy and momentum 
\[
H\equiv -P_t=\int dx\, T^t_t(t,x)\;,~~~P_x=-\int dx\, T^t_x(t,x)\;,
\]
commute with each other and define the evolution of local operators in
time and space.

{\bf 2.}{\it Positivity of energy.} The Hilbert space of the
theory is spanned by the eigenvectors of $H$ and
$P_x$; the eigenvalues of $H$ are non-negative with $H=0$ only for
the vacuum.

{\bf 3.}{\it Unitarity.} The theory possesses a Hilbert space with
positive-definite norm. The EMT is Hermitian.

{\bf 4.}{\it Existence of the dilatation current.} The symmetry
(\ref{txscaling}) corresponds to a current $D^\m$ which is
related to the EMT by \cite{Wess:1960,Polchinski:1987dy}
$D^\m=\xi^\n T_\n^\m+V^\m$, 
where $V^\m$ is a local operator. The dilatation charge
$D=\int dx \,D^t(t,x)$
has the standard commutation relations with the energy and momentum,
\be
\label{DPstand}
-i[D,P_\n]=P_\n\;.
\ee

{\bf 5.}{\it Discrete, positive, diagonalizable spectrum of dimensions.} 
The set of all local operators is
spanned by operators $\phi_i$ transforming in the canonical way under dilatations,
\be
\label{canonical}
-i[D,\phi_i(\xi)]=\D_i\phi_i(\xi)+\xi^\m\d_\m\phi_i(\xi)\;,
\ee
where $\D_i\geq 0$ form a discrete set. We also assume that the number
of operators of a given dimension $\D$ is finite.\footnote{In
  fact, it is sufficient to make this assumption only for the level
  $\D=2$ containing the (improved) EMT.}  

{\bf 6.}{\it Finite velocity of signal propagation.} 
All commutators of local operators vanish
outside a certain causal cone,
\be
\label{caus}
[\phi_i(t,x),\phi_j(0,0)]=0~,~~\text{if}~~|x|> v_{max} |t|.
\ee

Then the (improved) EMT
of the theory decomposes into a sum of local dimension-2 tensors 
\be
\label{nonint}
T^\m_\n=\sum_{a=1}^N T^{(a)\,\m}_{~~~\n}\;,
\ee  
which
are traceless, conserved, have linearly related off-diagonal
components and 
mutually commute: 
\bseq
\label{propts}
\begin{align}
\label{tri}
&T^{(a)\,\m}_{~~~\m}=0\;,~~~~~~~~~
\d_\m T^{(a)\,\m}_{~~~\n}=0\;,~~~~\\
\label{symmi}
&T^{(a)\,x}_{~~~t}+v^2_{a}\, T^{(a)\,t}_{~~~x}=0~,~~~v_a^2>0\;,\\
\label{commi}
&[T^{(a)\,\m}_{~~~\n},T^{(b)\,\l}_{~~~\r}]=0~~~~~~\text{for}~a\neq b\;.
\end{align}
\eseq

An immediate consequence of this result is the existence of $N$
conserved currents
\begin{align}
&m^{(a)\,t}=v_a^{-1}x\,T^{(a)\,t}_{~~~t}+v_at\,T^{(a)\,t}_{~~~x}\;,\notag\\
&m^{(a)\,x}=v_a^{-1}x\,T^{(a)\,x}_{~~~t}+v_at\,T^{(a)\,x}_{~~~x}\notag
\end{align}
generating Lorentz boosts with the ``speeds of light''
$v_a$. Moreover, 
$T^{(a)\,\m}_{~~~\n}$ give rise to
$N$ copies of the conformal algebra ${\cal C}_{a}$ whose generators commute for
$a\neq b$. Thus the symmetry of the theory is 
${\cal C}_{1}\otimes {\cal C}_{2} \otimes\ldots\otimes
{\cal C}_{N}$. In the case $N=1$ we obtain a
standard chiral CFT, while a parity symmetric CFT corresponds to
$N=2$, $v_1^2=v_2^2$. 

Note that if $N>1$ the theory contains sectors
which do not interact with each other. Indeed, consider the set 
$\{\Phi^{(a)}\}$ composed 
 of all local
operators that transform non-trivially under ${\cal C}_{a}$ and are
singlets with respect to ${\cal C}_{b}$, $b\neq a$ (this
set is not empty as it contains at least
$T^{(a)\m}_{~~~\n}$). The operators $P^{(a)}_\n=-\int dx\,
  T^{(a)t}_{~~~\n}$ act as translations on the operators
  $\{\Phi^{(a)}\}$ without affecting the operators of the family 
$\{\Phi^{(b)}\}$. This implies that the operators of the $a$-th family
cannot appear in the operator product expansion of 
$\{\Phi^{(b)}\}$ and vice versa. Hence the operator algebras
$\{\Phi^{(a)}\}$ and $\{\Phi^{(b)}\}$ are completely decoupled and
form two independent CFT's.

I now turn to the proof of (\ref{nonint}) which consists of 3 steps.

{\bf Step 1.} Let us make the EMT traceless by an appropriate
improvement. 
The conservation of $D^\m$ implies
$T^\m_\m+\d_\m V^\m=0$.
On the other hand, the EMT can be always supplemented by a total
divergence,
\be
\label{improve}
T^\m_\n\mapsto \hat{T}^\m_\n=T^\m_\n+\d_\l Y^{\l\m}_\n\;,
\ee 
where $Y^{\l\m}_\n$ is anti-symmetric in its upper indices. The
choice
$Y^{\l\m}_\n=\varepsilon^{\l\m}\varepsilon_{\n\r}V^\r$,
where $\varepsilon^{\l\m}$, $\varepsilon_{\n\r}$ are anti-symmetric
symbols, $\varepsilon^{tx}=-\varepsilon_{tx}=1$, 
yields $\hat{T}^\m_\m=0$ as desired. In what follows we omit the hats
on the improved EMT. 

A further improvement casts the scaling transformation of the EMT 
into the canonical form
(\ref{canonical}) with $\D=2$. To see this note that the commutation
relations (\ref{DPstand}) imply 
\be
\label{DT1}
-i[D,T^\m_\n]=2T^\m_\n+\xi^\l\d_\l T^\m_\n+A^\m_\n\;,
\ee 
where the local tensor $A^\m_\n$ is 
such that the integral $\int d\xi^\l\,\varepsilon_{\l\m} A^\m_\n$
vanishes for any contour interpolating between the spatial infinities. 
Then it must have 
the form 
$A^\m_\n=\varepsilon^{\m\l}\d_\l B_\n$
for some local $B_\n$.
The trace-free property of the EMT implies 
\be
\label{Bcurl}
\varepsilon^{\m\l}\d_\l B_\m=0\;.
\ee
Now we expand $B_\n$ in the complete basis defined in
(\ref{canonical}),  
\be
\label{Bexpans}
B_\n=\sum_i \beta_{\n,i}\phi_i.
\ee 
One notices that operators with dimension $\D_i=1$ cannot appear in
this sum. Indeed, expanding the EMT as
$T^\m_\n=\sum_i \gamma_{\n,i}^\m \phi_i$
and substituting into (\ref{DT1}) we obtain
\[
\sum_i(\D_i-2) \gamma_{\n,i}^\m \phi_i = \sum_i 
\beta_{\n,i}\epsilon^{\m\l}\d_\l\phi_i\;.
\]
Clearly, the l.h.s. does not contain operators of dimension 2. On the
other hand, the derivative of a dimension 1 operator would produce a term of
dimension 2 on the r.h.s., which would lead to contradiction. 
Then a redefinition of the form (\ref{improve}) with
\be
\label{EMTnew}
Y^{\l\m}_\n=
\varepsilon^{\l\m}\sum_i(\D_i-1)^{-1}\beta_{\n,i}\;\phi_i\;
\ee
brings the commutator of the EMT with $D$ to the desired
form. Note that the new EMT remains traceless:
substituting (\ref{Bexpans}) into (\ref{Bcurl}) we obtain
$\sum_i\epsilon^{\m\l}\beta_{\n,i}\d_\l\phi_i=0$.
The cancellation must hold within each subspace of operators with a
given dimension, which
translates into the trace-free property of the improvement term 
(\ref{EMTnew}).

{\bf Step 2.} This is the key step in the proof. We will demonstrate
the following property of the Fourier 
transformed EMT:\footnote{This is trivially valid for a $(1+1)$-dimensional
  relativistic theory where 
the EMT decomposes into the left- and right-moving parts,
$T^\m_\n(t,x)=T^{(l)\,\m}_{~~~\n}(t+x)+T^{(r)\,\m}_{~~~\n}(t-x)$,
and hence its Fourier transform is localized on the light-cone.}
\be
\label{EMTFourier}
\tilde T^\m_\n(\omega,k)\!\equiv\!\!\!\int \!\! dtdx\, 
\e^{-i\omega t+ikx}\, T^\m_\n(t,x)\!=\!0
~~~\text{for}~~ |k/\omega|\!<\!v_{max}^{-1}\;
\ee
First, we show that the Fourier components (\ref{EMTFourier})
annihilate the vacuum,
\be
\label{vacannihl}
\tilde T^\m_\n(\omega,k)\ket{0}=0~~~~\text{for}~~ |k/\omega|<v_{max}^{-1}\;.
\ee
Consider the expectation value of the commutator of two identical EMT
components (no summation over the
repeated indices !), 
\be
\label{spectr}
F^{~\m}_{0\,\n}\!(\xi)\!\!\equiv\!\!\bra{0}[T^\m_\n(0),T^\m_\n(\xi)]\ket{0}
\!=\!\!\!\int\!\! d\omega dk\,\e^{i\omega
  t-ikx}\!\rho^\m_\n(\omega,\! k)\! 
-\!\mathrm{h.c.,}
\ee
where the spectral density is defined as usual,
\be
\label{spectrdens}
\bra{0} \tilde T^\m_\n(\omega',k') \tilde T^\m_\n(\omega,k)\ket{0}
\!=\!(2\pi)^4\delta(\omega'+\omega)\delta(k'+k)\, \rho^{\m}_{\n}(\omega,k).
\ee
The positivity of energy and scale invariance fix
the form of the spectral density,
$\rho^\m_\n(\omega,k)=\theta(\omega)\,\omega^2{\hat\rho}^\m_\n(k/\omega)$.
The conservation and the
trace-free property of the EMT yield the relations between the
spectral functions corresponding to the different components,
\be
\label{densrel}
{\hat\rho}^t_x(\vk)=\vk^2{\hat\rho}^x_x(\vk)
=\vk^2{\hat\rho}^t_t(\vk)=\vk^4{\hat\rho}^x_t(\vk)\;.
\ee
Changing the integration 
variable in (\ref{spectr}) we obtain,
\be
\begin{split}
\label{spectr1}
F_{0}(t,x)&=\int d\omega d\varkappa
\theta(\omega)\,\omega^3{\hat\rho}(\varkappa)\,
\e^{i\omega(t-\varkappa x)}-\mathrm{h.c.}\\
&=2\pi i \int d\varkappa\,
{\hat\rho}(\varkappa)\,\delta'''(t-\vk x)\\
&=\frac{2\pi i}{x^3|x|}{\hat\rho}'''(t/x)\;,
\end{split}
\ee
where for clarity
we suppressed the indices $\m,\n$.
Then (\ref{caus}) implies that 
${\hat\rho}'''(\vk)$ vanishes if $|\vk|<v_{max}^{-1}$ and hence
${\hat\rho}(\vk)$ is at most quadratic
in its argument. 
Comparing to (\ref{densrel}) we conclude that the spectral densities
of all EMT components vanish at $|k/\omega|<v_{max}^{-1}$, implying  
(\ref{vacannihl}) by unitarity.

To extend Eq.~(\ref{vacannihl}) to any state in the Hilbert space let
us replace (\ref{spectr}) by the expectation value in a thermal 
ensemble with temperature $T$ (again, no summation over
the repeated indices~!),
 \be
\label{spectrtemp}
F^{~\m}_{T\,\n}(\xi)\!\!\equiv\!\!\langle[T^\m_\n(0),T^\m_\n(\xi)]\rangle_T
\!\!=\!\!\!\int \!\! d\omega dk\,\e^{i\omega t-ikx}\!
\rho^{~~\m}_{T\,\n}(\omega,\! k)\!-\!\mathrm{h.c.,}
\ee
where the thermal spectral density $\rho^{~~\m}_{T\,\n}(\omega,k)$ 
is defined by the relation (\ref{spectrdens}) with the vacuum
correlator on the l.h.s. replaced by the thermal average. 
In a unitary theory it is positive definite and vanishes if and only
if $\tilde T^\m_\n(\omega,k)=0$. 
Due to scale invariance the spectral densities 
entering (\ref{spectrtemp}) have the form\footnote{We again omit 
the indices $\m,\n$.}
$\rho_T(\omega,k)=\omega^2\hat\rho_1\big(\omega/T,k/\omega\big)$.
The dependence of $\hat\rho_1$ on the first argument prevents us from
explicitly taking the integral over frequency, so we have to take a
different route. Let us subtract from the thermal
average of the commutator its vacuum value,
\be
\label{Delta}
\begin{split}
{\cal D}(t,x;T)&\equiv F_T(t,x;T)-F_0(t,x)\\
=T^4&\int d\hat\omega d\hat k\;\hat\omega^2\;
\e^{-i Tx (\hat k-\hat\omega t/x)}\\
&\times\!\!\Big[(1-\e^{-\hat\omega})\hat\rho_1\big(\hat\omega,\hat k/\hat\omega\big)
-\epsilon(\hat\omega)\,\hat\rho\big(\hat k/\hat\omega\big)\Big]\;,
\end{split}
\ee
where 
$\epsilon(\hat\omega)$ is the sign function. 
In deriving (\ref{Delta})
we have rescaled the integration variables and
 used the
standard connection between the thermal spectral densities for
positive and negative 
frequencies,\footnote{Note
 that $\rho_T(\omega,k)$ is related to the imaginary part of the
retarded Green's function in the thermal state:
$2\Im G_T^{ret}(\omega,k)=(1-\e^{-\omega/T})\rho_T(\omega,k)$.}
$\rho_T(-\omega,-k)=\e^{-\omega/T}\rho_T(\omega,k)$.
We want to consider
(\ref{Delta}) as a function of the variables $x$ and 
$\theta\equiv t/x$. Due to (\ref{caus}) 
this function vanishes for $|\theta|<v_{max}^{-1}$ and non-zero $x$ 
at any $T$. However, the behavior at 
$x=0$ is subtle: ${\cal D}$ 
is a distribution in the two-dimensional space
$(t,x)$ and its restriction to a line $t=\theta x$ must be appropriately
specified. 
In the appendix we show that 
\be
\label{lemma}
x^3 {\cal D}(\theta x,x;T)=0~~~\text{for}~~|\theta|<v_{max}^{-1}\;
\ee
in the well-defined sense of one-dimensional distributions 
in the variable $x$.
Then multiplying 
(\ref{Delta}) by $x^3$ and taking the integral over $x$ we obtain,
\be
\label{zerodens1}
\frac{\d^3}{\d \theta^3}\int_0^\infty \frac{d\hat\omega}{\hat\omega}\,
(1-\e^{-\hat\omega})\hat\rho_1(\hat\omega,\theta)=0\,,\;\text{if}~|\theta|<v_{max}^{-1}\;,
\ee  
where we have used that the integrand is an even function of
$\hat\omega$ and, as proved above, the vacuum spectral density
vanishes 
for these
values of $\theta$. Next we use the 
relations (\ref{densrel}) which remain
valid in the thermal bath. Combined with (\ref{zerodens1}) 
they imply $\rho^{~~\m}_{T\,\n}(\omega,k)=0$ for
$|k/\omega|<v_{max}^{-1}$ and 
 (\ref{EMTFourier}) follows by unitarity. 

{\bf Step 3.}
We now construct a sequence
of
local dimension-2 operators $\Phi_n(\xi)$, such that
\bseq
\begin{align}
\label{Phisn}
&\d_t\Phi_n=\d_x\Phi_{n+1}\;,\\
\label{Phis1}
&\Phi_1=-T^t_x~,~~~~\Phi_2=T^x_x=-T^t_t~,~~~~\Phi_3=T^x_t\;.
\end{align}
\eseq
Consider the integral over space of the EMT component $T^x_t$. Its
average over a normalizable state is
\[
\bra{\alpha}\!\!\int\! dx\, T^x_t(t,x)\ket{\alpha}
\!=\!\int\frac{d\omega}{2\pi}\e^{i\omega t} 
\bra{\alpha}\tilde T^x_t(\omega,k=0)\ket{\alpha}.
\]
Due to (\ref{EMTFourier}) the matrix element on the r.h.s.
is proportional\footnote{It cannot contain derivatives of the
  $\delta$-function as in the vicinity of $k=0$ the matrix element 
$\bra{\alpha}\tilde T^x_t(\omega,k)\ket{\alpha}$ is a product of a
positive distribution $\bra{\alpha}\tilde T^t_t(\omega,k)\ket{\alpha}$
and a bounded function $(\omega/k)\Theta\big(v_{max}-|\omega/k|\big)$.} 
to $\delta(\omega)$
implying that
$\int dx\; T^x_t(t,x)$ is time independent.
Hence there exists a local operator $\Phi_4$ such that
$\d_t T^x_t=\d_x \Phi_4$. Its Fourier transform is 
$\tilde\Phi_4(\omega,k)=(\omega/k)\,
\tilde T^x_t(\omega,k)$,
and thus also satisfies the property (\ref{EMTFourier}). Then by
applying
to $\Phi_4$
the same reasoning as above one concludes that there must
exist another operator $\Phi_5$ such that $\d_t\Phi_4=\d_x\Phi_5$ and
so on.

According to the assumption~5, the number $N$ of linearly independent
operators in the sequence $\Phi_n$ is finite. 
Inside the linear envelope of the operators
$\Phi_n$, $n\leq N$, we can define a map 
$\Phi\mapsto\tilde\Phi$, where $\d_t\Phi=\d_x\tilde\Phi$. It is
straightforward to show that this map is Hermitian with respect to the scalar
product
\[
\langle \Phi,\Psi\rangle\!=\!\int d\xi_1 d\xi_2\;f^*(\xi_1) f(\xi_2)\;
\bra{0}\Phi^+(\xi_1)\Psi(\xi_2)\ket{0}\;,
\]
where $f(\xi)$ is an arbitrary test function introduced to make the
integral finite. 
Therefore, it
can be diagonalized by a linear transformation,
\be
\label{diagon}
\Phi_n=\sum_{a}c_{na}\Psi_a~,~~~~
\d_t\Psi_a=v_a\d_x\Psi_a\;,
\ee
where $\Psi_a$ are local operators and the eigenvalues $v_a$ are
real. 
This yields the decomposition (\ref{nonint}) of the EMT with\footnote{If
  one imposes invariance under 
 spatial parity it is possible to show that the eigenvalues
$v_a$ come in sign-symmetric pairs $\pm v_a$. Then the partial EMT's
(\ref{decomp}) can be grouped into EMT's of parity-symmetric subsystems.} 
\be
\label{decomp}
\begin{split}
&T^{(a)\,t}_{~~~~x}=-c_{1a}\Psi_a~,~~~~
T^{(a)\,x}_{~~~~x}=-T^{(a)\,t}_{~~~~t} =c_{2a}\Psi_a\;,\\
&T^{(a)\,x}_{~~~~t}=c_{3a}\Psi_a\;.
\end{split}
\ee
Eqs.~(\ref{Phisn}), (\ref{diagon}) imply the relations
between the coefficients $c_{n+1,a}\!\!=\!\!v_a c_{na}$ which yield the
properties (\ref{tri}), (\ref{symmi}).

The property (\ref{commi}) will follow if we prove that the operators
$\Psi_a$, $\Psi_b$ corresponding to different
eigenvalues $v_a\neq v_b$ commute at the coincident
time which we 
choose to be $t=0$. 
Indeed, this will imply that the time-evolution of the
operators for different $a$ is completely independent, being given by
the 
Hamiltonians
$H^{(a)}\propto \int dx \Psi_a$, and thus
the vanishing of the commutator is preserved for a non-zero temporal
separation.
Locality fixes the most general form of the equal-time commutator,
\be
\label{comgen}
[\Psi_a(0,x),\Psi_b(0,0)]=\sum_{i=0}^3 Q_i(0,x)\delta^{(i)}(x)
\ee
where $\delta^{(i)}(x)$ is the $i$th derivative of the
$\delta$-functions and the local operators $Q_i(t,x)$ have dimensions $3-i$. 
The sum terminates because we have
assumed the spectrum of dimensions to be non-negative. 

Now
consider the Jacobi identity
\[
-i[P_t,[\Psi_a,\Psi_b]]=-i[\Psi_a,[P_t,\Psi_b]]
+i[\Psi_b,[P_t,\Psi_a]]\;.
\]
From (\ref{comgen}) one reads the l.h.s.
\be
\label{timeder}
\sum_{i=0}^3\dot Q_i(0,x)\delta^{(i)}(x)\;.
\ee
While on the r.h.s. upon substituting the time derivatives of
$\Psi_{a,b}$ by their space derivatives using (\ref{diagon}) we find a
term proportional to the fourth derivative of the $\delta$-function, 
$(v_a-v_b)Q_3(0,x)\delta^{(4)}(x)$. This must vanish implying $Q_3=0$. 
But then
the contribution with $\delta^{(3)}$ disappears from (\ref{timeder}) and
requiring that on the r.h.s. it also vanishes we obtain $Q_2=0$.
Continuing this argument we conclude that
the commutator vanishes altogether.

This completes the proof. $\blacksquare$

\paragraph{Appendix: analysis of ${\cal D}(t,x;T)$}

Scale invariance and continuity in temperature imply the 
relations,
\bseq
\begin{align}
\label{rescaling}
&{\cal D}(\l x,\l t;T/\l)=\l^{-4}{\cal D}(x,t;T)\;,\\
\label{limit}
&\lim_{T\to 0}{\cal D}(t,x;T)=0\;.
\end{align}
\eseq
These must be understood in the sense of two-dimensional
distributions.
We define the object
\be
\label{obj}
\begin{split}
{\cal D}(\theta x,x;T)&=\int dt\,\delta(t-\theta x)\, {\cal D}(t,x;T)\\
=\lim_{\l\to\infty}&\int dt\, \l\varphi\big(\l(t-\theta x)\big)\,
{\cal D}(t,x;T)\;,
\end{split}
\ee
where we have regularized the last expression  
by approximating the $\delta$-function
with a sequence of smooth functions of compact
support. Consider the convolution of (\ref{obj})
with an arbitrary test function weighted by $x^3$
\be
\label{lims}
\begin{split}
&\int dx\,x^3 {\cal D}(\theta x,x;T) f(x)\\
&=\lim_{\l\to\infty}\int dxdt\, x^3\varphi(t-\theta x)\,f(x/\l)\, 
{\cal D}(t,x;T/\l)\;.
\end{split}
\ee
Vanishing of ${\cal D}$ at $|x|>v_{max}|t|$ implies that if
$|\theta|<v_{max}^{-1}$ the function $f(x/\l)$ in the integrand can be
replaced by $f_1(x;\l)\equiv\varphi_1(x)f(x/\l)$ where $\varphi_1(x)$ is a
localized test function that does not depend on $\l$ 
and equals 1 in a certain interval around
$x=0$ (this interval is determined by the overlap of the causal cone
with the support of $\varphi(t-\theta
x)$). The sequence $f_1(x;\l)$ has a well-defined limit
$\varphi_1(x)f(0)$ at $\l\to\infty$. Then (\ref{limit}) implies that
(\ref{lims}) vanishes leading to
(\ref{lemma}).

\paragraph{Acknowledgments} I have greatly profited from the 
discussions with Diego Blas, Dmitry Gorbunov, Sergei Dubovsky, Charles
Melby-Thompson, Dmitry Levkov, Emin Nugaev, Oriol Pujolas, Slava Rychkov and Misha
Shaposhnikov. I thank Fedor Bezrukov and Valery Rubakov for valuable comments
on the first version of the paper.
This work was supported in part by 
the Grant of the President of Russian Federation NS-2835.2014.2.


\begin{thebibliography}{99}

\bibitem{Zamolodchikov:1986gt} 
  A.~B.~Zamolodchikov,
  JETP Lett.\  {\bf 43}, 730 (1986)
  [Pisma Zh.\ Eksp.\ Teor.\ Fiz.\  {\bf 43}, 565 (1986)].

\bibitem{Polchinski:1987dy} 
  J.~Polchinski,
  Nucl.\ Phys.\ B {\bf 303}, 226 (1988).

\bibitem{Luty:2012ww} 
  M.~A.~Luty, J.~Polchinski and R.~Rattazzi,
  JHEP {\bf 1301}, 152 (2013)
  [arXiv:1204.5221 [hep-th]].

\bibitem{Dymarsky:2013pqa} 
  A.~Dymarsky, Z.~Komargodski, A.~Schwimmer and S.~Theisen,
  ``On Scale and Conformal Invariance in Four Dimensions,''
  arXiv:1309.2921 [hep-th].

\bibitem{Nielsen:1978is} 
  H.~B.~Nielsen and M.~Ninomiya,
  Nucl.\ Phys.\ B {\bf 141}, 153 (1978).

\bibitem{Chadha:1982qq} 
  S.~Chadha and H.~B.~Nielsen,
  Nucl.\ Phys.\ B {\bf 217}, 125 (1983).

\bibitem{Iengo:2009ix} 
  R.~Iengo, J.~G.~Russo and M.~Serone,
  JHEP {\bf 0911}, 020 (2009)
  [arXiv:0906.3477 [hep-th]].

\bibitem{Giudice:2010zb} 
  G.~F.~Giudice, M.~Raidal and A.~Strumia,
  Phys.\ Lett.\ B {\bf 690}, 272 (2010)
  [arXiv:1003.2364 [hep-ph]].

\bibitem{Anber:2011xf} 
  M.~M.~Anber and J.~F.~Donoghue,
  Phys.\ Rev.\ D {\bf 83}, 105027 (2011)
  [arXiv:1102.0789 [hep-th]].

\bibitem{Vafek:2002jf} 
  O.~Vafek, Z.~Tesanovic and M.~Franz,
  Phys.\ Rev.\ Lett.\  {\bf 89}, 157003 (2002)
  [cond-mat/0203047].

\bibitem{Franz:2002qy} 
  M.~Franz, Z.~Tesanovic and O.~Vafek,
  Phys.\ Rev.\ B {\bf 66}, 054535 (2002)
  [cond-mat/0203333].

\bibitem{Lee:2002qza} 
  D.~J.~Lee and I.~F.~Herbut,
  Phys.\ Rev.\ B {\bf 66}, 094512 (2002)
  [cond-mat/0201088].

\bibitem{Herbut:2009qb} 
  I.~F.~Herbut, V.~Juricic and B.~Roy,
  Phys.\ Rev.\ B {\bf 79}, 085116 (2009)
  [arXiv:0811.0610 [cond-mat.str-el]].

\bibitem{Giuliani:2011dc} 
  A.~Giuliani, V.~Mastropietro and M.~Porta,
  Annals Phys.\  {\bf 327}, 461 (2012)
  [arXiv:1107.4741 [cond-mat.str-el]].

\bibitem{Grover:2013rc} 
  T.~Grover, D.~N.~Sheng and A.~Vishwanath,
  Science, \ {\bf 344}, 280 (2014)
  [arXiv:1301.7449 [cond-mat.str-el]].

\bibitem{Sundrum:2011ic} 
  R.~Sundrum,
  Phys.\ Rev.\ D {\bf 86}, 085025 (2012)
  [arXiv:1106.4501 [hep-th]].

\bibitem{Bednik:2013nxa} 
  G.~Bednik, O.~Pujol\`as and S.~Sibiryakov,
  JHEP {\bf 1311}, 064 (2013)
  [arXiv:1305.0011 [hep-th]].

\bibitem{Hofman:2011zj} 
  D.~M.~Hofman and A.~Strominger,
  Phys.\ Rev.\ Lett.\  {\bf 107}, 161601 (2011)
  [arXiv:1107.2917 [hep-th]].

\bibitem{Wess:1960}
J.~Wess, 
Nuovo\ Cim.\ {\bf 18}, 1086 (1960).

\end{thebibliography}
\end{document}